\documentclass[twocolumn,showpacs,superscriptaddress,amsmath,amssymb]{revtex4}

\usepackage{graphicx}
\usepackage{dcolumn}
\usepackage{bm}

\begin{document}

\title{Evidence for Oxygen Holes due to $d$-$p$ Rehybridization in 
Thermoelectric Sr$_{1-x}$Rh$_2$O$_4$}

\author{Y.~Ishida}
\author{T.~Baba}
\author{R.~Eguchi}
\author{M.~Matsunami}
\author{M.~Taguchi}
\author{A.~Chainani}
\affiliation{RIKEN SPring-8 Center, Sayo, Sayo, Hyogo 679-5148, Japan}

\author{Y.~Senba}
\author{H.~Ohashi}
\affiliation{JASRI/SPring-8, Sayo, Sayo, Hyogo 679-5198, Japan}

\author{Y.~Okamoto}
\affiliation{ISSP, University of Tokyo, Kashiwa-no-ha, Kashiwa, 
Chiba 277-8561, Japan}

\author{H.~Takagi}
\affiliation{RIKEN, The Institute for Physical and 
Chemical Research, Wako, Saitama 351-0198, Japan}

\author{S.~Shin}
\affiliation{RIKEN SPring-8 Center, Sayo, Sayo, Hyogo 679-5148, Japan}
\affiliation{ISSP, University of Tokyo, Kashiwa-no-ha, Kashiwa, 
Chiba 277-8561, Japan}

\date{\today}

\begin{abstract}

Soft-x-ray photoemission and absorption spectroscopies are 
employed to investigate the electronic structures of 
Sr$_{1-x}$Rh$_2$O$_4$. Similarly to the 
layered cobaltates such as Na$_{1-x}$CoO$_2$, 
a valence-band satellite feature (VBS) 
occurs at higher binding energy to the O 2$p$ band. 
We find that the VBS resonates at the O 1$s$ edge. 
Additionally, core absorption 
shows clear $x$ dependence in the O 1$s$ edge 
rather than in the Rh 3$p$ edge. 
These results indicate that the holes in the initial state 
mainly have O 2$p$ character presumably due to 
$d$-$p$ rehybridizations affected by Sr$^{2+}$ vacancy potentials. 
The resultant inhomogenous charge texture may have 
impact on the TE transport properties at low $x$. 
\end{abstract}


\maketitle

The search for efficient thermoelectric (TE) materials 
is extensively pursued with the aim at practical applications 
such as TE batteries and Peltier refrigerators \cite{Mahan_Today}. 
Since 
metallic materials had been considered to exhibit poor TE performance 
\cite{Mahan_Today}, 
it was a surprise that a low-resistive layered cobaltate 
Na$_{1-x}$CoO$_2$ ($x$\,$\textless$\,0.5) 
exhibited large TE power ($Q$) at high temperatures 
\cite{Terasaki, NMat_HighNa}. 
The cobaltates have gained further interest as exhibiting rich 
phase diagram \cite{MLFoo} including superconductivity \cite{Super} 
and 3D magnetism \cite{Sugiyama_PRB, Bayrakci}. 
The largeness of $Q$ has 
been discussed from a 
band-theoretical viewpoint \cite{Singh, Takeuchi, Singh2007, Kuroki_JPSJ} 
or from a correlated viewpoint 
\cite{Koshibae_PRB, Ong_Nature, PALee, Haerter}, 
or from a viewpoint that 
there is a coherent-to-incoherent crossover in the low-energy excitations 
with increasing $T$ 
\cite{Limelette, Ishida_JPSJ}. 
Furthermore, interesting Na orderings \cite{Patterning, NSRRC} 
that affect the electronic properties 
\cite{Marianetti_PRL07_Na, NaRichPotential, Julien} have been reported, 
but their impact on the TE properties is not clear at present. 

\begin{figure}[htb]
\begin{center}
\includegraphics{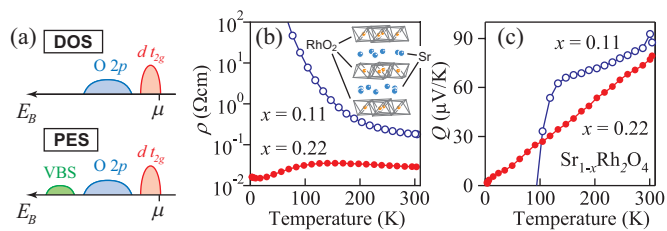}
\caption{\label{fig1} 
Sr$_{1-x}$Rh$_2$O$_4$, an analogue of Na$_{1-x}$CoO$_2$. 
(a) Electronic structure. The band theoretical DOSs of the
cobaltates and the rhodates exhibit $t_{2g}$ and O 2$p$ bands, but
the valence-band PES spectra additionally exhibit VBSs. 
Resistivity (b) and TE power (c) of 
Sr$_{1-x}$Rh$_2$O$_4$ 
as functions of temperature. 
The crystal structure of SrRh$_2$O$_4$ 
is shown in the inset of (b). 
}
\end{center}
\end{figure}

From a band-theoretical viewpoint, the valence band 
of NaCoO$_2$ 
consists of a filled $t_{2g}$ band positioned just below the 
chemical potential ($\mu$) and an O 2$p$ band 
at higher binding energy ($E_B$) as schematically shown in 
the upper panel of Fig.\ \ref{fig1}(a). 
With Na$^+$ deintercalaion, $\mu$ is shifted into the $t_{2g}$ band, and 
holes of mainly $t_{2g}$ character are introduced into the 
triangular lattice of Co$^{3+}$ ions. 
Valence-band spectra of Na$_{0.7}$CoO$_2$ recorded by 
photoemission spectroscopy (PES)
indeed show the $t_{2g}$ and the O 2$p$ bands, 
but in addition, there is a valence-band satellite 
feature (VBS) at $E_B$\,$\sim$\,11\,eV \cite{Hasan_PRL04} 
as schematically shown in the lower panel of 
Fig.\ \ref{fig1}(a). 
Similar VBSs occur in 
other layered cobaltates such as 
LiCoO$_2$ \cite{vanElp_LCO} and Ca$_3$Co$_4$O$_9$ \cite{Takeuchi} 
(VBSs in TE Bi-Sr-Co-O system are obscured by 
Bi 6$s$ states at $E_B$\,$\sim$\,11\,eV \cite{Mizokawa}). 
Thus, the VBS is a ubiquitous feature in the 
TE cobaltates, which is missing in the band-theoretical 
density of states (DOS).

Herein, we investigate element-specific 
electronic structures of 
Sr$_{1-x}$Rh$_2$O$_4$ \cite{Okamoto} 
using soft-x-ray absorption (XAS) and resonant PES. 
Sr$_{1-x}$Rh$_2$O$_4$ is structurally and 
electronically analogous to Na$_{1-x}$CoO$_2$: 
hole carriers are introduced into the layered triangular lattice of 
low-spin Rh ions (nominally $t_{2g}^6$) through Sr$^{2+}$ 
deintercalation [inset of Fig.\ \ref{fig1}(b)] to show 
insulator-to-metal transition 
at $x$\,$\sim$\,0.2 [Fig.\ \ref{fig1}(b)], 
and the $x$\,=\,0.22 sample shows 
$Q\,\sim$\,70\,$\mu$V/K at 300\,K 
[Fig.\ \ref{fig1}(c)]\cite{Okamoto}. 
Similarly to the cobaltates \cite{Hasan_PRL04, Takeuchi, vanElp_LCO}, 
we find a VBS in Sr$_{1-x}$Rh$_2$O$_4$. 
Moreover, the VBS resonates at the O 1$s$ edge 
followed by O 1$s$2$p$2$p$ Auger emissions, 
providing strong constraints on its origin. 
We also find clear $x$ dependence 
in the O 1$s$ XAS rather than in the Rh 3$p$ XAS. 
The results indicate that holes in Sr$_{1-x}$Rh$_2$O$_4$  
have strong O 2$p$ character presumably due to 
so-called $d$-$p$ rehybridizations 
\cite{Zunger, Marianetti_NMat04, Marianetti_PRL04, Zunger_Nature} 
that redistribute the holes from $d$ states to $p$ states 
beyond a rigid-band-shift picture.

Single-phase well-sintered Sr$_{1-x}$Rh$_2$O$_4$ ($x$\,=\,0.11 and 0.22) 
were prepared by a 
conventional solid state reaction as described elsewhere \cite{Okamoto}. 
The resistivity ($\rho$) and $Q$ 
of the samples [Fig.\ \ref{fig1}(b) and (c), respectively] 
nicely reproduced those reported previously \cite{Okamoto}. 
XAS and PES were performed at BL17SU of SPring-8 equipped with 
a VG Scienta SES2002 analyzer \cite{Ohashi}. 
Sample surfaces were obtained by fracturing the samples inside 
the spectrometer under ultrahigh vacuum ($<5\times$10$^{-8}$\,Pa). 
XAS spectra were recorded at 300\,K 
in the total electron yield method. 
PES spectra were recorded at 50\, K 
at $\sim$250-meV energy resolution and $E_B$ was referenced 
to $\mu$ of Au in contact with the sample and the analyzer. 
PES spectra were normalized to the incident photon flux. 

\begin{figure}[htb]
\begin{center}
\includegraphics{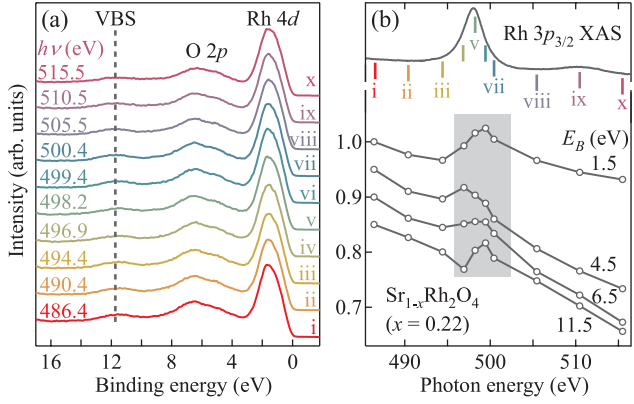}
\caption{\label{fig2} Rh 3$p$\,-\,4$d$ resonant PES of 
Sr$_{1-x}$Rh$_2$O$_4$ ($x$\,=\,0.22). 
(a) Valence-band spectra recorded across the Rh 3$p_{3/2}$ edge. 
(b) CIS spectra and Rh 3$p_{3/2}$ XAS. CIS spectra are normalized 
to the intensity at 
486.4\,eV and have arbitrary offsets. 
The labels (i\,-\,x) on the spectra in (b) 
correspond to the photon energies indicated by bars on the  
Rh 3$p$$_{3/2}$ XAS in (c). }
\end{center}
\end{figure}

Figure \ref{fig2}(a) shows valence-band spectra of 
Sr$_{1-x}$Rh$_2$O$_4$ ($x$\,=\,0.22) recorded across the Rh 3$p_{3/2}$ edge. 
We find a VBS at $E_B$\,$\sim$\,11.7\,eV, 
a feature missing in the LDA DOS \cite{WS_LDA}, as well as the 
Rh 4$d$ and the O 2$p$ bands 
centered at $E_{\rm B}$\,$\sim$\,1.5 
and 6\,eV, respectively. 
Figure \ref{fig2}(b) shows constant-initial-state (CIS) spectra at 
$E_B$\,=\,1.5, 4.5, 6.5, and 11.5\,eV. 
One can see resonant enhancement of Rh 4$d$ states 
at the Rh 3$p_{3/2}$ edge [shaded area in Fig.\ \ref{fig2}(b)]
in all features including the VBS (CIS at $E_B$\,=\,11.5\,eV). 
This indicates that the Rh 4$d$ weight is spread over a wide energy range 
and that the VBS has some Rh 4$d$ character. 
The presence of the Co 3$d$ character in the VBSs of the cobaltates 
was similarly confirmed through 
Co 3$d$ resonant PES \cite{Takeuchi, Hasan_PRL04, vanElp_LCO}.

\begin{figure}[htb]
\begin{center}
\includegraphics{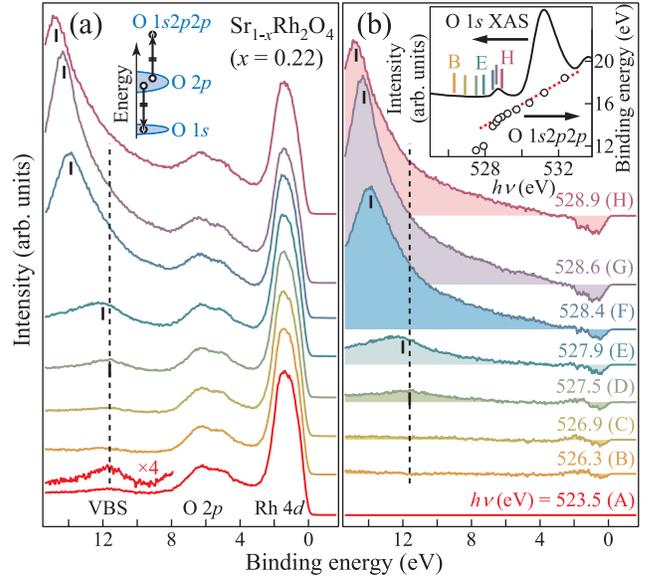}
\caption{\label{fig3} Valence-band spectra of 
Sr$_{1-x}$Rh$_2$O$_4$ ($x$\,=\,0.22) recorded across the 
O 1$s$ edge. 
(a) Valence-band spectra recorded in the vicinity of the O 1$s$ 
edge. Inset shows a schematic of the final 
state of an O 1$s$2$p$2$p$ 
Auger-electron emission. 
(b) Difference spectra to the 523.0-eV spectrum. 
The labels (A\,-\,H) on the spectra 
correspond to the photon energies indicated on the 
O 1$s$ XAS in the inset in (b). 
The dotted vertical lines indicate the 11.7-eV 
VBS, and the bars indicate 
the O 1$s$2$p$2$p$ Auger 
peaks. The O 1$s$2$p$2$p$ Auger peak positions are plotted in the 
inset in (b), in which the dotted line indicates the 
constant kinetic energy. }
\end{center}
\end{figure}

Next, 
we performed resonant PES at the O 1$s$ edge to obtain information 
about the O 2$p$ states. 
Figure \ref{fig3}(a) and (b) show, respectively, 
the valence-band spectra recorded in the vicinity of the O 1$s$ edge and 
the difference to the off-resonant spectra recorded 
at $h\nu$\,=\,523.0\,eV. 
One can see that the VBS resonates at $h\nu$\,=\,527.5\,eV, 
and subsequently, O 1$s$2$p$2$p$ Auger 
peak emerges from the vicinity of the 
VBS. 
The results indicate that the final state of the 11.7-eV VBS is 
similar to that of the O 1$s$2$p$2$p$ Auger emission, namely, 
the O 2$p$-two-hole final state [see schematic in Fig.\ \ref{fig3}(a)]. 
This interpretation is 
the same as that of the 6-eV satellite of Ni 
identified to a Ni 3$d$-two-hole final state \cite{Ni6eV}. 
In order to reach the O 2$p$ two-hole final state 
by photoemission, 
the initial state 
should contain an electronic configuration 
that has a single hole in the O 2$p$ state. 
We hereafter denote the initial oxygen-hole configuration as $p^5_{\rm v}$. 
We note that the $p^5_{\rm v}$ state is different from the 
ligand-hole states of the configuration-interaction 
CoO$_6$ cluster-model analyses \cite{Wu_XAS, KrollSawatzky}, 
since the $p_{\rm v}^5$ state is considered to be 
affected by the 
cation vacancy potentials \cite{footnote1} 
(discussed later). 
At $h\nu\ge$ 528.7\,eV, the O 1$s$2$p$2$p$ Auger peak 
position is shifted to higher $E_B$ 
since the kinetic energy of an O 1$s$2$p$2$p$ Auger electron
is independent of $h\nu$ 
[inset in Fig.\ \ref{fig3}(c)]. 
The resonant peak position at 
$h\nu$\,=\,527.5\,eV slightly deviates from the 
constant kinetic energy of the normal O 1$s$2$p$2$p$ Auger, 
perhaps since the $p_{\rm v}^5$ configuration is 
mixed to some $d$-hole configurations as 
inferred from the Rh 4$d$ resonant PES 
[Fig.\ \ref{fig2}].

\begin{figure}[htb]
\begin{center}
\includegraphics{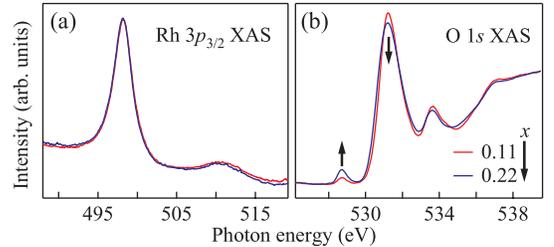}
\caption{\label{fig4} Rh 3$p_{3/2}$ (a) and O 1$s$ (b) XAS. 
The arrows in (b) indicate the change with increasing $x$. }
\end{center}
\end{figure}

Evidence for oxygen holes in the initial state is also found in the 
XAS as shown in Fig.\ \ref{fig4}. With increasing $x$, 
the height of the prepeak 
feature in the O 1$s$ XAS at 
$h\nu$\,=\,528.7\,eV becomes large and that of the main peak 
at $h\nu$\,=\,531.2\,eV becomes small (for each composition, 
the line shape at $h\nu$\,$\textless$\,533\,eV was reproducible for 
3 fractures). 
On the other hand, the Rh 3$p$ XAS is hardly changed with $x$. 
Since the 
O 1$s$ prepeak intensity is almost in proportional to $x$, 
we assign it to Sr-vacancy induced states mainly having O 2$p$ character, 
so that the prepeak is assigned to 
(1$s$)$^2$(2$p^5_{\rm v}$)\,$\to$\,(1$s$)$^1$(2$p^6$). 
The spectral-weight transfer seen in O 1$s$ XAS is often 
taken as a signature of strong electron correlation \cite{Wu_XAS}. 
In the XAS studies of (Li/Na)$_{1-x}$CoO$_2$ 
\cite{LCO_XAS_Abbate, LCO_XAS_Galakhov, Kroll_XAS}, 
main changes with $x$ 
occurred in the O 1$s$ XAS rather than in the 
Co 2$p$ XAS. 
Thus, oxygen holes in the initial state are common features in the 
TE rhodates and cobaltates.

The oxygen holes in Sr$_{1-x}$Rh$_2$O$_4$ 
revealed by resonant PES and XAS at the O 1$s$ edge 
indicate that the electronic-structure evolution with $x$ 
goes beyond a rigid-band-shift picture. 
Otherwise, $\mu$ 
would shift with increasing $x$ into the Rh 4$d$ band 
resulting in holes that mainly have Rh 4$d$ 
character. 
The spectral weight should therefore be redistributed with 
increasing $x$, most likely due to $d$-$p$ rehybridization 
\cite{Zunger, Marianetti_NMat04, Marianetti_PRL04, Zunger_Nature}: 
as evidenced from the charge densities 
calculated \cite{footnote2} 
for Li$_{1-x}$CoO$_2$ \cite{Zunger, Marianetti_NMat04} 
and Na$_{1-x}$CoO$_2$ \cite{Marianetti_PRL04}, 
the holes introduced via Li/Na deintercalation are not 
homogenously distributed in the Co layers but 
reside at oxygen sites neighboring the 
Li/Na vacancies, 
i.e., the cation vacancy potential binds the holes to form the 
$p^5_{\rm v}$ state 
\cite{Marianetti_NMat04}. 
The charge density at a Co site was nearly 
unchanged with increasing $x$ 
\cite{Zunger, Marianetti_NMat04, Marianetti_PRL04}, 
since the $t_{2g}$ holes were dressed by 
$e_{g}$ electrons transferred from the O 2$p$ states through 
the $d$-$p$ rehybridization. 
The essence of the charge rearrangement with doping can be captured 
in a simple model, namely, transition-metal impurities in semiconductors 
\cite{Haldane, Zunger_Nature}. 
The rigidity of the Rh 3$p$ XAS line shape with $x$ 
[Fig.\ \ref{fig4}(a)] is thus considered as a fingerprint 
that the $d$-$p$ rehybridization is self-regulating the 
local charge density about the Rh site to a nearly constant value.

A striking difference between the O 1$s$ XAS of Sr$_{1-x}$Rh$_2$O$_4$ 
and those of (Na/Li)$_{1-x}$CoO$_2$ 
\cite{LCO_XAS_Abbate, LCO_XAS_Galakhov, Wu_XAS, Kroll_XAS} 
is that the $x$-dependent prepeak in the former 
is well separated from the main peak [Fig.\ \ref{fig4}(b)], 
whereas those in the latter are merging 
into the main peaks. 
This can be understood that the degree of 
localization of the $p^5_{\rm v}$ state is affected by the 
strength of the cation vacancy potentials. 
Since the divalent Sr$^{2+}$ vacancy potential is 
stronger than those of the monovalent Li$^{+}$/Na$^{+}$, 
the holes are more strongly bound around the vacancies 
in the former than in the latter \cite{Marianetti_NMat04}. 
Hence, the $p^5_{\rm v}$ state appears to be more localized in 
Sr$_{1-x}$Rh$_2$O$_4$, so that the prepeak appears to 
be a sharp level. 
The localized character of the $p_{\rm v}^5$ state 
is 
supported by an $^{17}$O NMR study of Na$_{1-x}$CoO$_2$ ($x$\,=\,0.28), 
revealing that $\sim$30\,\% of the oxygen sites carry 
local magnetic moments \cite{Imai}. 
We naturally identify these magnetic oxygens 
to have the $p_{\rm v}^5$ configuration.

It should be noted that a VBS was also observed in stoichiometric LiCoO$_2$ 
having no Li vacancies \cite{vanElp_LCO}. Through a 
configuration-interaction cluster-model analysis, 
the VBS of LiCoO$_2$ was attributed to 
large $d$-$p$ hybridization and orbital degeneracy of the $d$ states 
\cite{vanElp_LCO}. 
Thus, the $d$-$p$ rehybridization with doping 
effectively occurs when the 
parent compound has large $d$-$p$ hybridization and orbital degeneracy. 
Large $d$-$p$ hybridization generally occurs in $t_{2g}$ electron 
system having unoccupied $e_{g}$ orbitals (this includes 
$d^0$ insulators \cite{Okada, Shin}). In fact, 
$d$-$p$ rehybridization was reported to occur 
when carriers are doped into SrTiO$_3$, 
a nominally $d^0$ insulator \cite{Ishida}. 
It is also interesting that a 
layered Cu$_x$TiSe$_2$, which shows 
TE properties as good as the cobaltates \cite{CuTiSe2}, 
is also considered to exhibit $d$-$p$ rehybridization \cite{LiTiS2}. 
Here, the hybridization of the Se 4$p$ states 
into the unoccupied Ti 3$d$ states opens the channel of 
$d$-$p$ rehybridization, and Cu$^{+}$ act as a source of 
``occupancy'' potential. 
Thus, the TE cobaltates, rhodates, and the intercalated Ti dichalcogenides 
can be categorized to those exhibiting $d$-$p$ rehybridization 
affected by the cation vacancy/occupancy potentials.

Since the holes in the initial state are not homogenously 
distributed in the Rh layers as expected in a rigid-band-shift 
picture but are further redistributed due to the 
$d$-$p$ rehybridizations and the Sr vacancy potentials, 
the charge density is considered to be nonperiodic compared to the 
crystallographic periodicity of SrRh$_2$O$_4$ 
(please see the nonperiodic charge densities calculated 
for (Li/Na)$_{1-x}$CoO$_2$ 
\cite{Zunger, Marianetti_NMat04, Marianetti_PRL04}.) 
Thus, it would be necessary to realize that the transport is 
occurring on such an inhomogenous charge texture 
with O 2$p$ holes bound to cation vacancies. 
For example, as was pointed out 
in \cite{WS_LDA}, the nonmetallic conduction 
at $x$\,$\lesssim$\,0.2 of Sr$_{1-x}$Rh$_2$O$_4$ \cite{Okamoto} 
can be viewed as a variable-range hopping, 
i.e., the low-energy excitations relevant 
to the transport show weak localizations as they are 
subject to randomness \cite{Mott}. 
The mobility-edge crossing occurring at $x$\,$\sim$\,0.2 in 
Sr$_{1-x}$Rh$_2$O$_4$ is reasonablly 
larger than that of Na$_{1-x}$CoO$_2$ 
occurring at $x$\,$\textless$\,0.1 \cite{NMat_HighNa}, 
since the vacancy potential of Sr$^{2+}$ is 
stronger than that of Na$^{+}$, although 
single crystal data will be helpful for further investigations 
\cite{Sugiura}. 
Another point to be noted is, when a nonmetallic transport is 
realized by randomness in a spin-orbitally degenerate system 
such as in Fe$_3$O$_{4-x}$F$_x$ \cite{Fe3O4}, 
a hump feature occurs at $T$\,$\sim$\,100\,K 
in a $Q$-$T$ curve
\cite{Fe3O4, Zvyagin, Sugiura}. 
This feature is very similar to the enhanced $Q$ at 
$\sim$100\,K in metallic Na$_{1-x}$CoO$_2$ at low $x$ \cite{NMat_HighNa}, 
which is considered to be on the verge of the metal-nonmetal transition. 
Further studies are necessary to clarify the relationship between 
the randomness and the transport properties in the 
rhodates and the cobaltates particularly at low $x$. 

In summary, we have performed resonant PES and XAS 
on Sr$_{1-x}$Rh$_2$O$_4$ 
and find that the holes have strong O 2$p$ character. 
The VBS, which commonly occurs in the TE cobaltates, 
is proven from resonant PES at the O 1$s$ edge to be 
a fingerprint of the O 2$p$ holes in the initial state. 
The results indicate a non-rigid-band 
evolution of the electronic structure with doping 
due to the $d$-$p$ rehybridization affected by the cation vacancy 
potentials 
\cite{Zunger, Marianetti_NMat04, Marianetti_PRL04, Zunger_Nature}, 
resulting in an inhomogenous charge texture that may affect 
the transport properties at low $x$.

Y.I.\ acknowledge A.~Mizutani, K.~Sugiura, H.~Ohta, and I.~Matsuda 
for informative discussion 
and T.~Mizokawa for critical 
reading of the manuscript.


\begin{thebibliography}{10}
\bibitem{footnote2}
Even though LDA DOS 
cannot reproduce VBS 
since LDA does not treat dynamical correlations in the 
electron removal state \cite{Kunes}, 
charge densities, 
which  are initial-state properties and are considered to 
be less affected by dynamical correlations, are 
nicely explained by LDAs.
\end{thebibliography}
\end{document}